\input harvmac

\let\includefigures=\iftrue
\let\useblackboard=\iftrue
\newfam\black

\includefigures
\message{If you do not have epsf.tex (to include figures),}
\message{change the option at the top of the tex file.}
\input epsf
\def\figin{\epsfcheck\figin}\def\figins{\epsfcheck\figins}
\def\epsfcheck{\ifx\epsfbox\UnDeFiNeD
\message{(NO epsf.tex, FIGURES WILL BE IGNORED)}
\gdef\figin##1{\vskip2in}\gdef\figins##1{\hskip.5in}
\else\message{(FIGURES WILL BE INCLUDED)}%
\gdef\figin##1{##1}\gdef\figins##1{##1}\fi}
\def\DefWarn#1{}
\def\figinsert{\goodbreak\midinsert}
\def\ifig#1#2#3{\DefWarn#1\xdef#1{fig.~\the\figno}
\writedef{#1\leftbracket fig.\noexpand~\the\figno}%
\figinsert\figin{\centerline{#3}}\medskip\centerline{\vbox{
\baselineskip12pt\advance\hsize by -1truein
\noindent\footnotefont{\bf Fig.~\the\figno:} #2}}
\bigskip\endinsert\global\advance\figno by1}
\else
\def\ifig#1#2#3{\xdef#1{fig.~\the\figno}
\writedef{#1\leftbracket fig.\noexpand~\the\figno}%
\global\advance\figno by1}
\fi
%

\useblackboard
\message{If you do not have msbm (blackboard bold) fonts,}
\message{change the option at the top of the tex file.}
\font\blackboard=msbm10 scaled \magstep1
\font\blackboards=msbm7
\font\blackboardss=msbm5
\textfont\black=\blackboard
\scriptfont\black=\blackboards
\scriptscriptfont\black=\blackboardss

\else

\fi
%
\def\yboxit#1#2{\vbox{\hrule height #1 \hbox{\vrule width #1
\vbox{#2}\vrule width #1 }\hrule height #1 }}
\def\fillbox#1{\hbox to #1{\vbox to #1{\vfil}\hfil}}
\def\ybox{{\lower 1.3pt \yboxit{0.4pt}{\fillbox{8pt}}\hskip-0.2pt}}
%
%


\def\comments#1{}



\def\II{\relax{I\kern-.10em I}}

\def\IZ{\relax\ifmmode\mathchoice
{\hbox{\cmss Z\kern-.4em Z}}{\hbox{\cmss Z\kern-.4em Z}}
{\lower.9pt\hbox{\cmsss Z\kern-.4em Z}}
{\lower1.2pt\hbox{\cmsss Z\kern-.4em Z}}
\else{\cmss Z\kern-.4emZ}\fi}
\def\IB{\relax{\rm I\kern-.18em B}}
\def\IC{{\relax\hbox{$\inbar\kern-.3em{\rm C}$}}}
\def\ID{\relax{\rm I\kern-.18em D}}
\def\IE{\relax{\rm I\kern-.18em E}}
\def\IF{\relax{\rm I\kern-.18em F}}
\def\IG{\relax\hbox{$\inbar\kern-.3em{\rm G}$}}
\def\IGa{\relax\hbox{${\rm I}\kern-.18em\Gamma$}}
\def\IH{\relax{\rm I\kern-.18em H}}
\def\II{\relax{\rm I\kern-.18em I}}
\def\IK{\relax{\rm I\kern-.18em K}}
\def\IP{\relax{\rm I\kern-.18em P}}

%

\def\inbar{\,\vrule height1.5ex width.4pt depth0pt}

\font\cmss=cmss10 
\def\IR{\relax{\rm I\kern-.18em R}}

%


%

\def\lp10{\ell_p^{10}}
\def\lp11{\ell_p^{11}}
\def\R11{R_{11}}

\def\frac#1#2{{#1 \over #2}}



\def\1dag{^{1\dagger}}
\def\2dag{^{2\dagger}}

\def\R#1#2#3{{{R_{#1}}^{#2}}_{#3}}



\hyphenation{Di-men-sion-al}



\lref\epr{A. Einstein, B. Podolsky and N. Rosen,
``Can quantum mechanical description of physical reality be considered complete?,''
{\it Phys. Rev.} {\bf 47} (1935) 777.}

\lref\waldgao{S. Gao and R. Wald, ``Theorems on gravitational 
time delay and related issues,''
{\it Class. Quant. Grav.} {\bf 17} (2000) 4999; gr-qc/0007021.}

\lref\sussbas{L. Susskind and N. Toumbas, ``Wilson Loops as Precursors,''
{\it Phys. Rev.}  {\bf D61} (2000) 044001; hep-th/9909013.}

\lref\toumbchinkski{J. Polchinski, L. Susskind and N. Toumbas,
``Negative Energy, Superluminosity and Holography,''
{\it Phys. Rev.} {\bf D60} (1999) 084006; hep-th/9903228.}

\lref\giddings{S. Giddings and S. Ross, 
``D3-brane shells to black branes on the Coulomb branch,''
{\it Phys. Rev.} {\bf D61} (2000) 024036; hep-th/9907204.}

\lref\lennyholography{
L.~Susskind,
``The World as a hologram,''
J.\ Math.\ Phys.\  {\bf 36}, 6377 (1995);
hep-th/9409089.
}

\lref\thooftholography{
C.~R.~Stephens, G.~'t Hooft and B.~F.~Whiting,
``Black hole evaporation without information loss,''
Class.\ Quant.\ Grav.\  {\bf 11}, 621 (1994);
gr-qc/9310006.
}
\lref\FreedmanGP{
D.~Z.~Freedman, S.~S.~Gubser, K.~Pilch and N.~P.~Warner,
``Renormalization group flows from holography supersymmetry and a c-theorem,''
Adv.\ Theor.\ Math.\ Phys.\  {\bf 3}, 363 (1999);
hep-th/9904017.
}
\lref\GiddingsPT{
S.~B.~Giddings and M.~Lippert,
``Precursors, black holes, and a locality bound,''
hep-th/0103231.
}

\lref\klebanetal{
M. Kleban, J. McGreevy, and S. Thomas,
``c-functions from black hole thermodynamics,''
to appear.}

\lref\israel{
W.~Israel,
``Singular Hypersurfaces And Thin Shells In General Relativity,''
Nuovo Cim.\ B {\bf 44S10}, 1 (1966)
[Erratum-ibid.\ B {\bf 48}, 463 (1966)].
}

\lref\adsreview{
O.~Aharony, S.~S.~Gubser, J.~Maldacena, H.~Ooguri and Y.~Oz,
``Large N field theories, string theory and gravity,''
Phys.\ Rept.\  {\bf 323}, 183 (2000)
[hep-th/9905111].
}

\lref\preskill{
D.~Beckman, D.~Gottesman, A.~Kitaev and J.~Preskill,
``Measurability of Wilson loop operators,''
arXiv:hep-th/0110205.
}

\lref\steve{
S. Shenker,
``What are strings made of?,'' 
talk at String Theory at the Millennium, Caltech, Jan. 15, 2000.
}

\Title{\vbox{\baselineskip12pt\hbox{hep-th/0112229}
\hbox{SU-ITP-01/38}}} 
{\vbox{ \centerline{Implications of bulk causality}
\bigskip
\centerline{for holography in AdS} }}
\bigskip
\bigskip
\centerline{Matthew Kleban, John McGreevy, and Scott Thomas}
\bigskip
\centerline{{\it Department of Physics, Stanford University,
Stanford, CA 94305}}
\bigskip
\bigskip
\noindent

Gravitational time delay in asymptotically Anti de Sitter spaces has
consequences for holographic duality. 
We argue that the requirement of bulk causality implies 
that it is not possible for a
collection of boundary observers, performing local measurements, to
extract information from precursors. Using similar arguments, 
we derive an integrated weak energy constraint on spacetimes 
which can admit a holographic dual. 

\bigskip

\Date{December, 2001}

\newsec{Introduction}

It is widely believed that quantum gravity is holographic,
in the sense that a
$d$-dimensional gravitational theory is dual to a
$(d-1)$-dimensional theory without a dynamical metric
\refs{\lennyholography,
\thooftholography}.
Several examples of such dualities have been discovered
\adsreview; we focus in this note on theories of the AdS/CFT type,
where gravity in an asymptotically anti-deSitter spacetime is dual
to a theory which is defined in the UV by a conformal gauge theory.

The conjecture of holographic duality is that the two theories
contain identical physical information.  However, their natural
variables are related through some very complex and non-linear
field redefinition, or perhaps by some large gauge fixing \steve, making
it very difficult in all but the simplest situations to 
decode the map directly. Nevertheless, one apparently inescapable
consequence of the conjecture is that any event occurring in the
bulk of the AdS space is {\it instantaneously} reflected in some
way in the boundary theory.  A good way to see this is to note
that when a localized bulk excitation arrives at the boundary, it
will excite local gauge invariant operators.  One can then evolve
this configuration backwards using the gauge theory Hamiltonian, 
leading to a state which cannot involve excited local
operators, and yet also cannot be the gauge theory vacuum.
Following \toumbchinkski \sussbas, we will refer to such a state
in the boundary theory reflecting a bulk event as a precursor,
because it exists on the boundary before any bulk gravitational
excitation could have propagated there. We will not attempt to
address the question of the precise nature of the precursor state.
Rather, we wish to ask if it can be detected in a manner which 
is consistent with bulk causality, and how a boundary 
theory containing any such effects can manage to be causal.

A candidate answer is that the information in the precursors is not
available to any single local observer, but instead is extended
spatially in a way roughly analogous to the famous {\it gedanken}
experiment proposed by Einstein, Podolsky, and Rosen \epr. In that
scenario, two spin measurements are performed at spacelike separated
points, and the results must be collected at some central location
before the surprising correlations can be detected. This
does not violate causality
because by the time the correlation in the results is reconstructed, the
future lightcones of the measurement events have intersected.

At first glance it appears that in a similar way one can avoid
causality violation in the case of AdS/CFT precursors. 
Consider a collection of spatially separated boundary observers, 
each able to 
perform a local measurement.  Having made their observations, these 
observers transmit their information along the boundary 
to a collection 
point at which precursor 
information about a bulk event might be reconstructed. 
One might imagine that the
collection time is such that a signal from the bulk will have
reached the boundary before any conclusions can be drawn by an
observer at the collection point, hence avoiding a violation of causality.

\ifig\compete{ A spatial section of AdS, showing the paths whose
transit times we compare. $1$ labels the path for a graviton
propagating from an event at the center of the bulk AdS space 
to the north pole of the
boundary sphere; $2$ indicates the path of a photon along the
boundary from the equator to the north pole. } 
{\epsfxsize2.0in \epsfbox{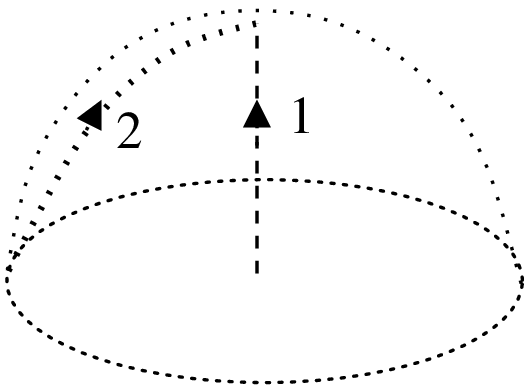}}
However, we will argue that such a resolution fails in
cases other than that of empty AdS space (dual to an exactly
conformal theory); the addition of matter to the bulk, which
breaks the conformal invariance of the boundary theory, would
spoil the effect and introduce causality violation. 
(because in this case the time it would take to assemble precursor 
information on the boundary
would be shorter than the time it takes for 
bulk signals to arrive at the boundary). 
Instead, we
will argue for an alternative explanation: the precursor
information can never be usefully measured 
by any set of correlated boundary observers with
access only to local gauge invariant quantities. Therefore, the
only indication of the existence of a precursor available to such
boundary observers will be {\it ex post facto}: after a bulk
signal has arrived on the boundary and excited a local operator, a
boundary physicist could use the time-reversed gauge theory
equations of motion to extrapolate backwards and deduce the
existence of the precursor.


In the next section we calculate transit times of massless
disturbances through the bulk and around the boundary of
asymptotically anti-de Sitter spacetimes.  We present a low-energy
approximation to a class of domain wall spacetimes which are
interesting in this regard. In \S3 we discuss the implications of
these calculations for precursors in the boundary theory, and in
\S4 we infer a constraint on which spacetimes of this type can have
a holographic dual.

\newsec{Time delay in AdS spacetimes}

First, we will review the causal structure of five-dimensional 
AdS space. The
metric in global coordinates is
\eqn\globalcoords{ ds^2 = - (1 + {r^2  \over R^2}) dt^2 + (1 +
{r^2 \over R^2})^{-1} dr^2 + r^2 d\Omega_3^2. } The spacetime can
be thought of as a cylinder over $S^3$, with time running down the
axis, $r = \infty$ as the boundary, and $r=0$ the center.  The time
for a massless particle to propagate along the boundary from the
north pole of the boundary sphere ($\theta = 0, r = \infty$) to
the equator ($\theta = \pi/2, r = \infty$) is
$$ t_{\rm BDY} = \lim_{r \to \infty} \int_0^{\pi \over 2} 
d\theta \sqrt{ g_{\theta \theta} \over g_{t t} } =
{\pi \over 2} \lim_{r \to \infty} \sqrt {r^2 \over 1 + {r^2 \over R^2}} =
{\pi R \over 2}.
$$
The time for a graviton to propagate from the center of the bulk 
spacetime ($r=0$) to
the boundary ($r = \infty$) is
$$ t_{\rm BULK} =
\int_0^{\infty} dr \sqrt{ g_{rr} \over g_{tt}} = \int_0^\infty {dr
\over 1 + {r^2 \over R^2}} = {\pi R \over 2} = t_{\rm BDY}.$$

Therefore, in pure AdS, a precursor reflecting an event at $r=0$
could in principle be measured instanteously by a
ring of boundary observers at the equator who then 
transmit their information along the boundary to the north pole. 
Since the time it takes for this information to arrive at 
the north pole 
is the same as the time it takes for bulk signals 
originating at $r=0$ to reach the north pole through the 
bulk, no violation of causality occurs, and there is no problem. 
As we show below, this is not the case if a domain wall exists in the 
bulk spacetime. 

\bigskip
\noindent
{ \it The Domain Wall Solution}

If the bulk spacetime is modified by including additional
stress-energy, the bulk path time is increased.  In order to
demonstrate this in a simple context, we will use a toy model for
the bulk five-dimensional effective gravitational theory. Consider gravity
coupled to a single scalar, with an asymmetric double well
potential for the scalar, admitting a kink solution.
\ifig\kink{
The scalar potential and profile in the domain wall model.
}
{\epsfxsize3.0in
\epsfbox{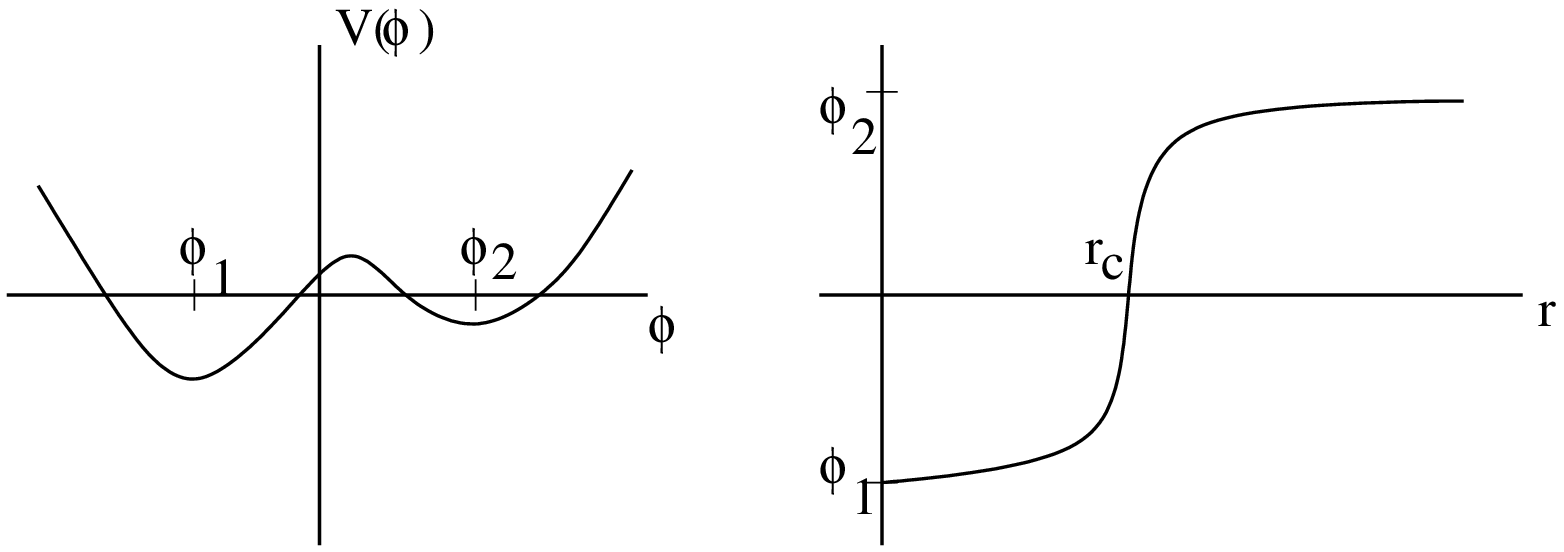}}
Away from the kink on either side, the space is AdS, with the
curvature determined by the value of the potential at the minimum
on that side. The kink itself is a domain wall, with tension
determined by the potential. This tension is balanced in
Einstein's equations by the difference in the AdS curvatures on
either side of the wall. Very similar solutions occur in string
theory examples where a particular mass deformation is added in
the boundary gauge theory, so that some reduced 
supersymmetry is preserved in the region interior to the domain
wall and the boundary 
theory flows in the IR to a conformal theory with a lower central
charge \FreedmanGP. The gravity description of this is essentially
the kink solution presented here.

We wish to find the metric for this domain wall spacetime.  We
will work in a limit where the wall is very thin, so that the
space is AdS everywhere except very near $r = r_c$, 
the position of the domain wall.  The most
general asymptotically AdS metric with ($S^3$) spherical symmetry
is
\eqn\adsmet{
ds^2 = -  \left( 1 +  {(r/r_0)^2 \over R^2} - {c \over r^2} \right)
(dt/t_0)^2  +
(r/r_0)^2 d\Omega^2_3 +
\left( 1 +  {(r/r_0)^2 \over R^2}  - {c \over r^2} \right)^{-1} (dr/r_0)^2 .}
Here $t_0$ and $r_0$ are simply coordinate rescalings, while $c$
is the coefficient of a Schwarzschild term. 

We have kept the coordinate
scalings explicit for the purpose of matching two such solutions
across a thin domain wall. For a domain wall with a delta function
tension, such a matching requires that the metric be continuous
(although its derivative will be discontinuous). We can always set
$t_0 = r_0 = 1$ outside the wall by an overall coordinate
redefinition. As mentioned above, the potential for the scalar
determines the value of the AdS radius $R$ on each side of the
wall, so the remaining parameters are $c$ inside and outside, and
$t_0$ and $r_0$ inside. The full metric is
\eqn\wallmet{\eqalign{
ds^2 =  & - \left( 1 + {r^2 \over R^2} - {c \over r^2} \right)
dt^2 + r^2 d\Omega_3^2 +
 \left( 1 +  {r^2 \over R^2} - {c \over r^2}
\right)^{-1} dr^2, \; \; (r_c < r < \infty) \cr ds^2  =  & -
\left( 1 + {(\rho/\rho_0)^2 \over R_-^2} \right) (dt/t_0)^2 +
 (\rho/\rho_0)^2 d\Omega^2_3  \cr &+
 \left( 1 +  {(\rho/\rho_0)^2 \over R_-^2}
\right)^{-1} (d\rho/\rho_0)^2 \; \; \;\;\;\;\; (\rho \equiv r -
r_c(1 - \rho_0), \;\;\; 0 < \rho < \rho_c \equiv r_c \rho_0).} }
We have set $c=0$ inside the wall which corresponds to pure
AdS space in the interior region. 
Notice that the domain wall has an ADM energy -
there is a Schwarzschild term in the exterior metric. 

Continuity of the metric requires $t_0 = 1/\rho_0$, and
$$ \rho_0^2 = (1 + r_c^2/R^2 - c/r^2)/(1 + r_c^2/R_-^2).$$ We also need
to impose Einstein's equations at the wall; this is most
conveniently done using the Israel \israel\ conditions \eqn\isr{
K^\alpha_\beta - \delta^\alpha_\beta K = \tau \delta^\alpha_\beta}
where $K_{\alpha \beta} = \nabla_\alpha n_\beta$, $n_\beta$ is a
unit normal vector, and $\tau$ is the tension of the domain wall.
This equation yields two independent equations for $\tau$:
\eqn\exein{ \tau = {3 \over r_c}  \left( \sqrt{1 + r_c^2/R_-^2} -
\sqrt{1 + r_c^2/R^2 - c/r_c^2}  \right)}  and \eqn\exeout{ \tau =
{3 \over (1 + r_c^2 / R_+^2 - c / r_c^2)^{1/2}} \left[ -
 r_c / R_+^2 -  c / r_c^3 +  \rho_c / R_-^2 \right].}
Equating these determines the ADM mass $c$ in terms of the
exterior and interior AdS radii $R$ and $R_-$, and the position of
the domain wall $r_c$.  An important consistency check is that the
Schwarzschild radius $r_h$ of the exterior metric satisfies $r_h <
r_c$ for all values of the parameters.  In other words, there is
no horizon anywhere in the spacetime, as expected.

Using this metric we can compute the bulk/boundary transit times
as above. The transit time along the boundary will be the same as
that of empty AdS space with the same asymptotic radius. The 
bulk transit time for a graviton travelling from the center to the
boundary of the domain wall spacetime can be computed exactly.  
However, it is somewhat unwieldy, so we
present an approximate expression valid for $r_c \gg R, R_-$:
\eqn\bulktime{ \eqalign{ t_{\rm bulk} &= \int_0^\infty dr \sqrt{
g_{rr} \over g_{tt} } = {t_0 \over \rho_0} \int_0^{\rho_c} { d\rho
\over 1 + { \rho^2 \over \rho_0^2 R_-^2} } + \int_{r_c}^\infty {dr
\over 1 + {r^2 \over R^2} - {c \over r^2}} \cr & \geq {\pi R \over
2} - R \tan^{-1} {r_c \over R} + {R_- \over \rho_0} \tan^{-1} {r_c
\over R_-} \simeq \pi R / 2 + {R \over r_c} \left (R - R_- \right)
+ O\left(R^3 \over r_c^3 \right) . }}

A more general static spherically symmetric solution can be built
up by a superposition of such domain walls, or by including other
types of matter.  For such configurations the bulk
transit time can be made arbitrarily long. 
This is clear from, for example, the fact that a
black hole horizon in the interior will send the time to infinity.
This corresponds to a finite temperature gauge theory,
where a low energy signal will be washed out by
thermal fluctuations.

While this work was in progress we became aware of \waldgao, in
which the authors prove a general theorem which demonstrates that
the bulk propagation time through as asymptotically AdS space is
always at least as great as that through an empty space with the same
asymptotics, as long as the weak energy condition is satisfied.

\newsec{Implications for precursors}

We are now in a position to discuss in more detail the question of
what sort of precursors can be allowed in a holographic theory. 
We are not concerned with the precise nature of the precursor, 
but rather with the circumstances under which it is in principle detectable.
Our results do not depend on the details of the precursor state.
To simplify the discussion, and as a concrete example, we will use
as a guide the proposal of \sussbas, in which the authors suggests
that the precursor will affect the expectation value of a spatial
Wilson loop in the boundary gauge theory. Roughly speaking, the
expectation value $\langle W \rangle$ is given by the area of the
minimal surface in bulk AdS bounded by the loop. If there is a
disturbance in the bulk at a depth such that the minimal surface
passes through it, then the authors of \sussbas\ argue $\langle W
\rangle$ will be modified. At first glance this appears to violate
bulk causality, as no local gauge invariant observables on the
boundary will be excited until the signal has had time to
propagate to the boundary. 
However, for pure AdS space
if the deepest point of the minimal surface bounding a circular Wilson loop 
just touches a bulk event, 
then a bulk signal originating at the event will reach 
the boundary  at exactly
the same time as a light signal sent from the edge of the loop
along the boundary to its center. This is consistent with the standard
intuition about the UV/IR correspondence.

There has recently been some debate in the literature over the
validity of this proposal \GiddingsPT; we will not attempt to
address this here.  As mentioned above, for our purposes the
precise nature of the precursor state is not important.  

To make the possible paradox in the domain wall spacetime 
as sharp as possible, consider an
event occurring in the center of the AdS space (meaning the origin
$r=0$ of the global coordinates \globalcoords).  To create such an
event we could send two particles from the north and south
poles of the boundary sphere, in such a way that they meet at
$r=0$ and form a small black hole, which then evaporates
isotropically at $t=0$.

The largest Wilson loop available (one whose associated Wilson
surface extends farthest into the bulk) is one that extends around the
equator of the boundary sphere at $r=\infty$. 
If the expectation
value of such a loop could be measured instantaneously 
by a set of local observers
spaced around the equator, each making instanteous local measurements, 
the information 
could be gathered by a boundary observer equi-distant from the 
equator, 
(by Santa Claus at the north pole, say) in time
$t = \pi R/2$, assuming the observers had massless boundary excitations at
their disposal with which to send the signal.  In pure AdS, this
is also the time for a graviton to propagate from $r=0$ to any
point on the boundary. However, as we have argued above, in a
bulk spacetime with additional matter such a signal will take longer to
arrive, $t > \pi R/2$.
Since the spacetime is still asymptotically AdS, the
field theory in the asymptotic UV is a CFT. 
In particular it contains
massless degrees of freedom with which boundary observers can 
send signals. 
Hence such measurements by Santa Claus to assemble precursor information 
from the collection of equatorial observers could
violate bulk causality, as he could become aware of the
bulk event via the boundary CFT measurements before a signal from 
the event could arrive directly 
through the bulk.

Perhaps in such a non-conformal theory the expectation value of a
Wilson loop will not detect the precursor configuration.  However
if there is any measurement a set of local observers can make, the
worst possible case would have them spread out over the entire
boundary sphere, rather than just the equator.  This at most
doubles the time it would take to collect information through 
massless boundary excitations: the
time is bounded by $\pi R$ (the path time for a boundary gluon to go from
South to north pole). However, the graviton path time through 
the bulk can be made
arbitrarily large.  
Therefore,

{{\narrower\smallskip\noindent
{\it A measurement of a generic precursor
state by any set of boundary observers with access only to local, gauge
invariant information is prohibited by bulk
causality.}\smallskip}}

While existence of precursors in the gauge theory is a 
necessary consequence of holographic duality, they are not
measurable by
any local observer or set of observers. The best that can be said
is that when, or if, a local gauge invariant operator is excited,
a boundary physicist could deduce via the equations of motion
that the precursor must have been present in the past.

This conclusion 
is consistent with the results of \preskill, which
demonstrated that 
an ordinary quantum 
mechanical measurement of a spatial Wilson loop 
in a nonabelian gauge theory is impossible.
The authors discuss what they term a ``destructive 
measurement,'' where a ring of observers measure 
small pieces of the loop by circulating charged matter.  
As noted by \preskill, this is not a measurement in the 
usual sense because it does not leave eigenstates 
undisturbed.

\newsec{Conclusions}

One interesting and important question that arises in discussions
of holography is whether or not gravity in a particular background
has a holographic dual.  An interesting condition on the class of
such spacetimes can be obtained by turning the above situation
around.\foot{We thank Joe Polchinski and Gary Horowitz for discussions on this
point.} Any asymptotically AdS spacetime that satisfies the weak
energy condition everywhere satisfies
\eqn\intwec{
    2 \int_{0} ^{\infty} d r
    \sqrt { g_{r r} \over g_{t t } } \geq \pi R
}

{However, if the weak energy condition is violated to a sufficient
extent that \intwec\ is {\it not} satisfied, a graviton could
travel from pole to pole through the bulk 
in a time less than $\pi R$.  Now imagine
in such a space an event at the South Pole of the boundary sphere
which sends energy in all directions into the bulk and along 
the boundary. 
Gravitons will reach Santa
Claus through the bulk before a gluon can arrive via the
boundary, and hence a local gauge invariant excitation will occur
at the north pole before the equations of motion of a causal boundary gauge
theory allow. Therefore,} 

{{\narrower\smallskip\noindent
{\it Any asymptotically AdS spacetime
must satisfy the integrated weak energy condition \intwec\ if it
is to have a causal holographic boundary dual}.\smallskip} } 

\noindent As an example, note that
in the domain wall solution equation \bulktime\ implies that a
negative tension wall (one with $R < R_-$) will violate this
condition and therefore can not have a causal field theory 
dual.\foot{Such domain walls would correspond to a field theory that
flows from fixed point to fixed point, but with $c_{UV} < c_{IR}$.
Therefore we see that for flows from CFT to CFT, the
four-dimensional c-theorem \FreedmanGP\ is equivalent to our
condition.}

One might imagine that by placing orientifolds or other consistent
negative-tension string theory objects in the interior of the
AdS space one can violate the weak energy condition while still maintaining
a stringy motivation for the existence of a holographic dual.
The integrated weak energy condition would still hold, however.

\bigskip

\centerline{\bf{Acknowledgements}}

We thank Ben Freivogel, Simeon Hellerman, Gary Horowitz, Shamit Kachru, 
Nemanja Kaloper, 
Joe Polchinski, and Lenny Susskind, for discussions.
This work was supported in part by the DOE under contract
DE-AC03-76SF00515 and by National Science
Foundation grant PHY00-97915.

\listrefs

\end